\documentclass[twocolumn,amsmath,amssymb,superscriptaddress]{revtex4}

\usepackage{graphicx} 
\usepackage{dcolumn}  
\usepackage{bm}       

\begin{document}

\title{Nonlocal interactions in doped cuprates: correlated motion of
Zhang--Rice polarons}

\author{L. Hozoi}
\affiliation{Max-Planck-Institut f\"{u}r Physik Complexer Systeme,
             N\"{o}thnitzer Str. 38, 01187 Dresden, Germany}
\affiliation{Max-Planck-Institut f\"{u}r Festk\"{o}rperforschung,
             Heisenbergstrasse 1, 70569 Stuttgart, Germany}

\author{S. Nishimoto}
\affiliation{Max-Planck-Institut f\"{u}r Physik Complexer Systeme,
             N\"{o}thnitzer Str. 38, 01187 Dresden, Germany}

\author{G. Kalosakas}
\affiliation{Max-Planck-Institut f\"{u}r Physik Complexer Systeme,
             N\"{o}thnitzer Str. 38, 01187 Dresden, Germany}

\author{D. B. Bodea}
\affiliation{Max-Planck-Institut f\"{u}r Physik Complexer Systeme,
             N\"{o}thnitzer Str. 38, 01187 Dresden, Germany}

\author{S. Burdin}
\affiliation{Max-Planck-Institut f\"{u}r Physik Complexer Systeme,
             N\"{o}thnitzer Str. 38, 01187 Dresden, Germany}

\date{\today}

\begin{abstract}
In-plane, inter-carrier correlations in hole doped cuprates are investigated by 
\textit{ab initio} multiconfiguration calculations.
The dressed carriers display features that are reminiscent of both Zhang--Rice
(ZR) CuO$_4$ singlet states and Jahn--Teller polarons. The interaction between 
these quasiparticles is repulsive.
At doping levels that are high enough, the interplay between long-range unscreened
Coulomb interactions and long-range phase coherence among the O-ion half-breathing
vibrations on the ZR plaquettes may lead to a strong reduction of the effective
adiabatic energy barrier associated to each polaronic state.
Tunneling effects cannot be neglected for a relatively flat, multi-well energy
landscape.
We suggest that the coherent, superconducting quantum state is the result of such
coherent quantum lattice fluctuations involving the in-plane O ions.
Our findings appear to support models where the superconductivity is related
to a lowering of the in-plane kinetic energy.
\end{abstract}


\maketitle

The parent compounds of the high-temperature cuprate superconductors are Mott
insulating antiferromagnets. Superconductivity (SC) occurs in these systems upon
hole or electron doping.
While the magnetic interactions are remarkably strong, there is increasing evidence
that many properties cannot be understood without taking into account the lattice
degrees of freedom.
Experiments that indicate strong electron--phonon (EP) couplings are the inelastic
neutron scattering \cite{CuO_INS_reznik06,CuO_INS_braden05},
angle-resolved photoemission (PE) \cite{CuO_PE_lanzara01,CuO_PE_gweon01},
X-ray absorption fine structure (XAFS) \cite{CuO_XAFS_acosta02}, and electron
paramagnetic resonance \cite{CuO_EPR_kochelaev97}.
Anomalies in the phonon, PE, and XAFS spectra were previously addressed
with $t$\,--$J$ or Hubbard-like models supplemented with EP interaction terms  
\cite{CuO_HHdmft_capone04,CuO_tJph_oliver04,CuO_tJph_ishihara04,CuO_tu_bishop03}.
Strong and anomalous electron--lattice couplings were also found by 
\textit{ab initio}, explicitly correlated calculations 
\cite{CuO_HNY_05,CuO_HN_05}.
In the case of hole doping \cite{note_RefUndConf}, multiconfiguration (MC)
calculations
on clusters of few CuO$_6$ octahedra show that the doped holes enter O $2p_x$ and
$2p_y$ orbitals that form $\sigma$-bonds with the open-shell Cu $3d_{x^2-y^2}$
orbitals \cite{CuO_HNY_05} and give rise to singlet states similar to the kind of
configuration proposed by Zhang and Rice (ZR) long time ago \cite{CuO_ZR_88}.
A major deviation from the original ZR picture is that the formation of such a
singlet is associated with significant lattice deformations.
The most stable configuration corresponds to Cu--O distances that are shorter by
5--6$\%$. 
Most remarkably, O-ion displacements that restore the translational symmetry 
induce strong charge redistribution.
The $2p$ hole, which for a distorted CuO$_4$ plaquette is equally distributed 
over the four anions, can be partially transferred onto a single ligand to give
an electronic wavefunction (WF) with a dominant contribution from a 
...--\,Cu\,$d^9$--\,O\,$p^5$--\,Cu\,$d^9$--... configuration.
The ZR-like singlet \textit{polaron} (ZRP) can hop thus within the
CuO$_2$ plane via such $d^9$--\,$p^5$--\,$d^9$ states through coupling to the
oxygen vibrations. The energy barrier associated with the hopping process is few
hundreds meV, for an isolated $2p$ hole \cite{CuO_HNY_05}.

$d^9$--\,$p^5$--\,$d^9$ entities were postulated to exist in cuprates by several
authors \cite{CuO_hizhny_88,CuO_emery88_1}.
This single-oxygen $2p$ hole and the adjacent Cu holes are viewed as a spin-1/2
quasiparticle state in which the $3d$ spins are ferromagnetically (FM) polarized,
i.e. coupled to a triplet, and the spin on the O ion is low-spin coupled to it.
Interactions that could lead to pairing of these quasiparticles and SC were
analyzed in \cite{CuO_hizhny_88,CuO_emery88_1} in terms of $p$\,--\,$d$, extended 
Hubbard Hamiltonians.
It was pointed out \cite{CuO_emery88_1,CuO_emery88_2} that such models imply
different low-energy physics as compared to the effective single-band model
proposed by ZR \cite{CuO_ZR_88}.
Interestingly, the \textit{ab initio} calculations \cite{CuO_HNY_05} show that
the two concepts, the ZR and the $d^9$--\,$p^5$--\,$d^9$ configurations, are
in fact intimately related. The connecting element is the lattice degree of
freedom, more precisely the Cu--O bond-length fluctuations.
Near-degeneracy effects between a quadrisinglet (QS) polaronic state similar
to the ZR state and a two-site spin-singlet polaron were studied in the frame of
a Holstein--Hubbard model in \cite{CuO_aubry_9900}.
Although some details are quite different, the two-site spin-singlet resembles
a two-hole $d^9$--\,$p^5$ singlet configuration.
It was shown that in the parameter region with near-degeneracy, the effective
mass of the QS polaron is sharply reduced.
Near-degeneracy plus quantum tunneling effects associated with a rather 
flat energy landscape are believed to be \cite{CuO_aubry_9900} crucial ingredients 
in the theory of ``bipolaronic'' SC.

In this Letter we investigate interactions between two or more spatially
separated ZRPs.
We find that long-range phase coherence among the O-ion half-breathing vibrations
on the ZR CuO$_4$ plaquettes may lead to a correlated motion of the $2p$ holes.
Several effects come here into play. The local electron--lattice couplings make
possible the transfer of the ZR singlet to an adjacent plaquette.
The longer-range inter-carrier interactions are responsible, at dopant 
concentrations that are high enough and in a ``dynamic'' regime, for a
significant lowering of the effective energy barrier seen by each ZR quasiparticle.
This is exactly the kind of picture proposed, in a somewhat different context,
by Hirsch \cite{CuO_color_hirsch9200}: single carriers are heavily dressed at very
low concentrations by the interaction with the local environment; they partially
``undress'' and their effective mass decreases at higher concentrations due to
collective inter-carrier interactions.

Our analysis is mainly based on results obtained by \textit{ab initio} electronic
structure methods from traditional quantum chemistry.
The calculations are performed on finite clusters. 
The many-electron WF is constructed as a full configuration-interaction
(CI) expansion in the space defined by a limited set of so-called active orbitals.
These are chosen as those orbitals that are expected to contribute to degeneracy
effects, that is, strong mixing between configurations which have the same, or
nearly the same, energy.
The rest of the orbitals are either doubly occupied in all configurations or empty.
The former set of orbitals is called inactive, the latter is referred to as the
virtual orbital set and spans the rest of the orbital space, defined from the
basis set used to build the ``molecular'' orbitals (MOs).
This is known as the Complete Active Space (CAS) MC approach 
\cite{QCbook_HJO_00}.
The variational parameters of the WF are \textit{both} the CI and MO coefficients. 

Inter-carrier interactions are studied by doping a 7-plaquette [Cu$_7$O$_{22}$]
linear cluster with two holes.
The length of the cluster corresponds roughly to the scale of the superconducting
coherence length.
We used the lattice parameters of the hole doped
La$_{1.85}$Sr$_{0.15}$CuO$_4$ compound \cite{LaCuO_xrd_cava87}, with the in-plane
lattice constant $a\!=\!3.78$\,\AA.
The 7-plaquette cluster is embedded in an array of formal ionic point charges
(PCs) at the experimental lattice positions that reproduce the crystal Madelung
potential.
The Cu$^{2+}$ and La$^{3+}$ nearest neighbors (n.n.) are represented by effective 
model potentials \cite{TIPs_CuSr}.
Trial calculations were first performed with embedding PCs that correspond to
a fully ionic picture of the undoped material, La$_2$CuO$_4$. In this situation,
due to the large mutual repulsion, the CAS Self-Consistent Field (SCF)
calculations converge to solutions where the $2p$ holes are each located on the
outermost CuO$_4$ plaquettes.
However, at not too small doping other $2p$ holes exist in the immediate vicinity.
In a next step, we decided to take into account the presence of other doped holes
by introducing few extra elementary positive charges around the cluster.
The results discussed in the following paragraphs were obtained with a PC 
embedding where eight positive charges were added. The positions of these
extra PCs are shown in Fig.\,1.
The reason behind choosing this type of arrangement is discussed below.

\begin{figure}[!t]
\includegraphics[angle=270,width=0.95\columnwidth]{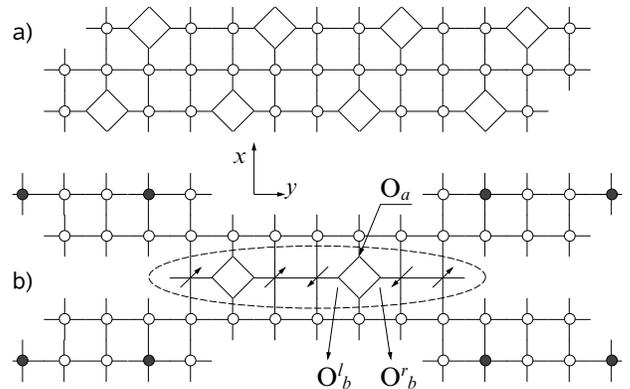}
\caption{ (a) 1/3-doped stripes at optimal doping in a ZR-like picture, see text. 
(b) The 7-plaquette cluster employed for the CAS\,SCF calculations and 
the positions of the extra positive charges (the filled circles) within the PC
embedding.
These correspond to the situation where every second pair of ZRPs along the
1/3-doped stripes is removed.
One of the two symmetry-equivalent lowest-energy configurations is sketched for
the ``quantum'' cluster.
Only the Cu sites are displayed. The squares represent distorted ZR plaquettes.}
\end{figure}

It is widely accepted \cite{CuO_stripes_kivelson03} that charge
and magnetic inhomogeneities appear in cuprates at least dynamically.
Goodenough \cite{CuO_jbg_review03} argued that at optimal doping
($x\!\approx\!1/6$) and low temperatures, 1/3-doped (fluctuating) stripes are
formed on every second chain of plaquettes.
In a ZR-like picture, the 1/3-doped stripes could be represented as in Fig.\,1(a),
for example.
The SC is associated in Ref.\cite{CuO_jbg_review03} with the
formation of traveling charge-density-waves \cite{CuO_cdw_gabovich01}
along the doped chains.
In models with EP coupling, the relevant phonon modes would be the O-ion
half-breathing modes with \textbf{q}\,=\,(1/3\,,\,0\,,\,0).
Experimental data that appear to support Goodenough's model are the strong
anomalies at such wave vectors, \textbf{q}\,=\,(0.25--0.33\,,\,0\,,\,0)
\cite{CuO_INS_reznik06,CuO_INS_braden05}. These anomalies are actually observed
in both hole and electron doped systems \cite{CuO_INS_braden05}.
Coming back to the positive charges added to our embedding, they would correspond
to the situation where every second pair of ZRPs in Fig.\,1(a) is removed,
i.e. $x\!=\!1/12$ doping.
In addition, we retain only the n.n. pairs around the ``quantum'' cluster.
For symmetry reasons, the extra positive charges on the left-hand side
are shifted by one plaquette to the right.

With the extra positive charge around the cluster, the two O holes are 
pushed from the outermost plaquettes toward the inner region.
To study electron--lattice couplings, we first determined the cluster geometry
configuration that minimizes the total energy. However, only the positions of those
ions are relaxed that have each neighbor represented at the all-electron level.
These are the Cu ions and the intervening ligands O$_b$.
Under the assumption of charge segregation into hole-rich and hole-poor regions
and filamentary conduction
\cite{CuO_hizhny_88,CuO_jbg_review03,CuO_stripes_kivelson03,CuO_tu_bishop03} 
the vibrations of these ions should be the most relevant.
The CAS\,SCF calculations were carried out with a minimal active space.
That is one orbital per hole, in our case nine electrons in nine orbitals.
It was shown \cite{CuO_HN_05} that the overall picture does not change when using
larger active spaces. Extra orbitals added to the active space turn out to have
occupation numbers either very close to 2.00 or to 0 \cite{CuO_HN_05}.
Also, when enlarging the active space, it is not obvious how many extra orbitals
are required for a balanced description of the different configurations.
All calculations were performed with the \textsc{molcas\,6} package
\cite{molcas6}.
\textit{All-electron} atomic natural-orbital basis sets from the \textsc{molcas}
library were employed, with the following contractions:
Cu $21s15p10d/5s4p3d$, O $14s9p/4s3p$.
As already mentioned, the apical oxygens are not included in the ``quantum''
cluster, but modeled by formal PCs.

\begin{table}[!t]
\caption{MPs illustrating the distribution of the O-holes.
The lowest-energy configuration corresponds to distorted, ZR plaquettes separated
by two other Cu $3d_{x^2-y^2}^{\,1}$ ions (1st column).
The ZR plaquettes are labeled with indices 1 and 2. Other notations are as in
Fig.\,1.
When the O$_b$ ions are shifted back to the middle positions, each hole is
partially transferred onto a single ligand (2nd column).
In parentheses, the same quantities are shown for a single doped hole, see text.}
\begin{ruledtabular}
\begin{tabular}{lll}
Configuration          &Dist. Cu--O$_b$\,,  &Undist. str.       \\
                       &ZRP states          &                   \\
\colrule
Cu$_1$ $3d_{x^2-y^2}$  &1.03 \ (1.03)       &1.09 \ (1.08)      \\
O$_{1b}^l$ $2p_y$      &1.64 \ (1.62)       &1.77 \ (1.76)      \\
O$_{1a}$ $2p_x$ (x2)   &1.61 \ (1.62)       &1.69 \ (1.68)      \\
O$_{1b}^r$ $2p_y$      &1.61 \ (1.62)       &1.29 \ (1.31)      \\
\\
Cu$_2$ $3d_{x^2-y^2}$  &1.03                &1.09               \\
O$_{2b}^l$ $2p_y$      &1.65                &1.77               \\
O$_{2a}$ $2p_x$ (x2)   &1.61                &1.68               \\
O$_{2b}^r$ $2p_y$      &1.60                &1.30               \\
\\
Rel. En. (meV)         &0.00 \ (0.00)       &650 \ (380)        \\
\end{tabular}
\end{ruledtabular}
\end{table}

The CAS\,SCF calculations always converge to states where the $2p$ holes are
separated by more than two lattice constants.
However, depending on geometry and the initial guess for the orbitals, these
states may have different character.
Two symmetry-equivalent \textit{minimum-energy} geometries were identified.
These are configurations where two ZRP states are formed on CuO$_4$
plaquettes that are separated by two other Cu ions, see Fig.\,1(b).
The spins on these intervening cations are antiferromagnetically (AFM) coupled.
For the ZR plaquettes, the Cu--O$_b$ distances are shorter by $5\%$.
Mulliken populations (MPs) of the relevant Cu $3d$ and O $2p$ atomic orbitals on
these plaquettes are listed in Table\,I (first column). 
The composition of the bonding (B) and antibonding (AB) $d_{x^2-y^2}$--\,$p_{x,y}$
$\sigma$-orbitals is depicted in Fig.\,2. Due to correlation effects, the 
occupation numbers (ONs) of the AB orbitals are relatively large.
Since the charge distribution on the two ZR plaquettes is nearly identical, only 
the orbitals on one of these plaquettes are shown in the figure.

When the bridging O$_b$ ions are shifted back to the middle positions
\cite{CuO_HNY_05}, each of the $2p$ holes is partially transferred onto a single
ligand.
This is illustrated in Table\,I and Fig.\,2. 
The results were obtained by static total energy calculations (adiabatic 
approximation).
It can be seen in Fig.\,2 that the doped-hole orbitals have now much weight onto
one of the adjacent plaquettes as well.
O-ion half-breathing displacements that shorten the Cu--O$_b$ bonds on these
\textit{adjacent} plaquettes induce propagation of the ZR singlet(s) along the 
Cu--O$_b$--... chain.
We found that the ZR $2p$ hole is already moved to the other plaquette for
distortions of less than $0.5\%$ from the high-symmetry geometry.
The adiabatic energy barrier is in the two-hole cluster 325\,meV per hole, see
Table\,I.
Separate calculations were performed with a single doped hole and no
extra positive charge in the PC embedding. The energy barrier is in this case
380\,meV \cite{note_OaRel}.
The energy difference between the two situations, 55\,meV per hole, is
considerable. In the presence of other \textit{mobile} $2p$ hole polarons in the
CuO$_2$ plane(s), the effect would be further enhanced and should lead to a strong
reduction of the polaron effective mass.
We arrive thus at Hirsch's paradigm \cite{CuO_color_hirsch9200,CuO_color_hirsch02}:
carriers are not happy being together, but they are happy \textit{moving together}.
This was described as a kinetic-energy-driven mechanism for SC 
and a change of color was predicted when entering the superconducting state
\cite{CuO_color_hirsch9200}. That the onset of SC is indeed
accompanied by a transfer of spectral weight from the visible region to the
infrared range was recently confirmed by Molegraaf \textit{et al.}
\cite{CuO_color_hajo02}.

\begin{figure}[!t]
\includegraphics[width=0.65\columnwidth]{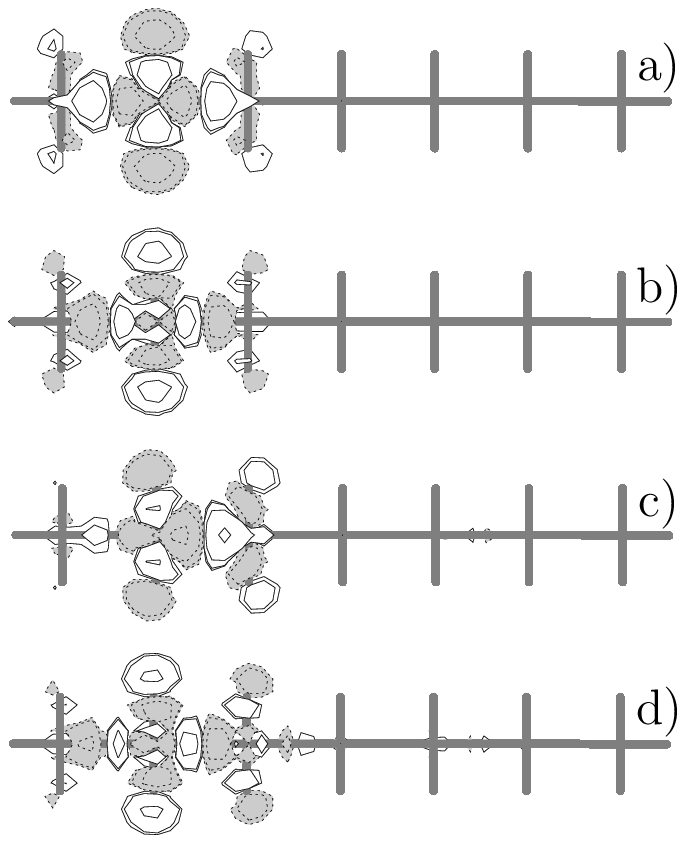}
\caption{(a)-(b) B and AB orbitals defining a ZR-like state on a distorted
plaquette. Their ONs are 1.86 and 0.14.
(c)-(d) B and AB orbitals in the undistorted structure.
The ONs are 1.74 and 0.26.
The $2p$ hole has now the largest weight onto a single ``chain'' oxygen,
see also Table\,I.
The whole 7-plaquette cluster is drawn.
The phases of the atomic orbitals are shown with different colors.}
\end{figure}

In comparison to the study from Ref.\cite{CuO_color_hirsch9200} and other model
Hamiltonian studies, we are able to identify much more precisely the crucial
interactions in the doped plane.
The CAS\,SCF results give also us hints about the nature of the pairing mechanism.
It is obvious that this is related to both long-range Coulomb repulsion effects
and shorter-range interactions.
Our results may be associated with the 4-plaquette correlation bag of Goodenough 
\cite{CuO_jbg_review03}.
The ZR singlet polarons in such a correlation bag are separated by AFM coupled
$d_{x^2-y^2}$ spins that bear some resemblance to the spin-singlet pairs of the
resonating-valence-bond state \cite{CuO_RVB_anderson87}.
Cu--O bond-length fluctuations may induce charge transfer along a chain of
plaquettes.
In the intermediate, undistorted geometry the quasilocalized $2p$ holes tend to
polarize FM the n.n. $3d$ spins.
In other words, the $3d$ spin on the adjacent plaquette is already ``prepared''
for the transfer of the $2p$ hole.
Also, strong antiferromagnetic correlations exist between the
$d^9$--\,$p^5$--\,$d^9$ units and the ``bipolaronic'' WF has a complex structure.
Our data suggest that two ZR-like quasiparticles could form a 4-plaquette 
bipolaronic singlet state only above a certain hole concentration. 
The distribution of ``doped'' holes displayed in Fig.\,1(b) corresponds (at the
scale of the figure) to 8--9$\%$ doping, only slightly higher than the 
concentration(s) where SC actually occurs. 
We also note that the effects described above, i.e. a lowering of the effective 
energy barrier due to longer-range interactions, are still obtained when
the positions of the extra positive embedding PCs are changed, individually or
collectively, by distances as large as one lattice constant.

More investigation is needed for a better understanding of these systems and 
quantitative estimates.
First, a more rigorous treatment of the dynamical electron correlation
\cite{QCbook_HJO_00} is required. 
Secondly, the electronic and lattice degrees of freedom should be described on
equal footing, beyond the adiabatic approximation.
Another non-trivial task is a self-consistent procedure to determine the (dynamic)
in-plane distribution of the charge carriers, perhaps in the spirit of the
dynamical mean-field theory \cite{note_ROHF}.
Nevertheless, the purpose of the present study is not to provide highly accurate
numbers, but to gain better insight into the very nature of the many-body 
in-plane interactions.
It is our belief that realistic models can only be constructed with the help 
of this type of first-principles investigations.

To conclude, using a WF-based \textit{ab initio} approach, we are able to describe
microscopic electron--lattice, magnetic, and nonlocal Coulomb interactions
characterizing the hole dynamics in (under)doped cuprates.
The dressed carriers display features which are reminiscent of both ZR spin-singlet
states and Jahn--Teller polarons.
We find that at doping levels that are large enough, the adiabatic energy
barrier associated to each polaronic state is significantly lower for the
synchronized, coherent ``hopping'' of two ZRPs than in the case of an isolated ZRP.
This is due to unscreened inter-carrier Coulomb interactions and requires phase
coherence among the O-ion half-breathing vibrations.
Tunneling effects cannot be neglected for a relatively flat, multi-well energy
landscape \cite{CuO_tu_bishop03,CuO_aubry_9900}.
Following the discussion from Ref.\cite{CuO_aubry_9900}, we suggest that the
coherent, superconducting phase is the result of such coherent quantum lattice
fluctuations involving the in-plane O ions.
Coherent, collective tunneling effects in the CuO$_2$ planes should lead to 
extended, ``resonant'' states \cite{CuO_aubry_9900} and a lowering of the in-plane
kinetic energy, see also
\cite{CuO_color_hirsch9200,CuO_color_hirsch02,CuO_KinEn_emery00,
CuO_colorDMFT_maier04}.
Polaronic behavior and local double-well potentials were also found in
the electron doped case \cite{CuO_HN_05}, which suggests some common ``global''
features for the two types of doping: quantum polarons coupled through spin
and longer-range unscreened Coulomb interactions.

\ 

We thank I. Eremin and P. Fulde for fruitful discussions.


\begin{thebibliography}{9}
\bibitem{note_OaRel} This value is somewhat larger than in \cite{CuO_HNY_05}
because the polarization of the apical-oxygen valence shells is neglected.
\bibitem{note_ROHF} A possibility is also periodic Restricted Open-Shell
Hartree--Fock calculations. Nevertheless, a unit cell of twelve plaquettes
at optimal doping, see Fig.1, is prohibitively large.
\end{thebibliography}
\end{document}